\documentclass{article}

\usepackage{amsmath,epsfig}

\def\ngs{\negthickspace \negthickspace \negthickspace \negthickspace}

\begin{document}

\title{Quark masses and weak couplings in the SM and
beyond}
\author{
   \\
   { Riccardo Barbieri}\\
   {\small\it Scuola Normale Superiore and INFN -- Sezione di Pisa}  \\
   {\small\it I-56100 Pisa, ITALIA}          \\
   {\small Internet: {\tt riccardo.barbieri@sns.it}}
}


\maketitle\abstract{I discuss  two topics: i) the
empirical adequacy of the Standard Model in the Flavour Sector in
view of recent data; ii) the possible existence of a hidden
structure in the quark masses and mixings based on textures.}

\section{Introduction}

I confess my embarrassment in trying to understand the precise
meaning of the title of this talk, as it was assigned to me by the
organizers. Needless to say, I could have asked them for
clarification. I realized however that, by not asking, I would
have been more free. Here is, therefore, my interpretation of what
the title means.

In the following three lines you see the Lagrangian of the Standard
Model (SM) in a concise but self-explanatory notation

\begin{align}
{\cal L}_{SM} &= \bar \Psi \not \! \! D \Psi + F_{\mu\nu}  F^{\mu\nu} \\
    &+ |D_{\mu} \varphi|^2 - V(\varphi) \\
    &+ \lambda \Psi \Psi \varphi
\end{align}

The three different lines correspond to the three different Sectors of
the SM: respectively the Gauge, the ElectroWeak Symmetry Breaking
(EWSB) and the Flavour Sectors (FS).

It is often said that the SM accounts for all data on the
fundamental interactions among elementary particles, except
gravity, in a satisfactory way. Although literally true, leaving
aside, for the moment, neutrino masses, a similar statement
ignores a strong asymmetry between the three Sectors in their
comparison with experiment. The Gauge Sector has passed in the
last decade a very severe scrutiny, which has brought also
evidence for its correctness in electroweak loop effects. The same
cannot simply be said for the EWSB Sector nor for the Flavour
Sector, the subject of this talk. Surprises are still possible or
even likely.

There are, in fact, two logically independent questions that are raised
by an examination of the FS of the SM:
$$\begin{cases}
\text{Is it empirically adequate?} \\
\text{Does it hide a deeper structure?}\\
\end{cases}$$
I shall try to address both these questions in the following. I
find it hard to tell which one of the two is more important, which
is not to say that the present understanding of the respectively
related problems is at comparable level of development.

\section{Comparison with present data}

\subsection{The predictions of the SM in the Flavour Sector}

As well known, the SM Lagrangian has a few sharp predictions in
the Flavour Sector, irrespective of the value taken by the
numerous parameters involved in the Yukawa couplings. They are:

\begin{description}
\item[In the Quark sector:]{All  flavour violations  and
  CP violation  reside in the weak charged current, with the amplitude
depicted in Fig.~1 \addtocounter{figure}{1}
being proportional to a unitary
matrix, the Cabibbo-Kobayashi-Maskawa (CKM) matrix $V_{ij}$.}

\begin{center}
\begin{minipage}[c]{.5\linewidth}
\centering \epsfig{file=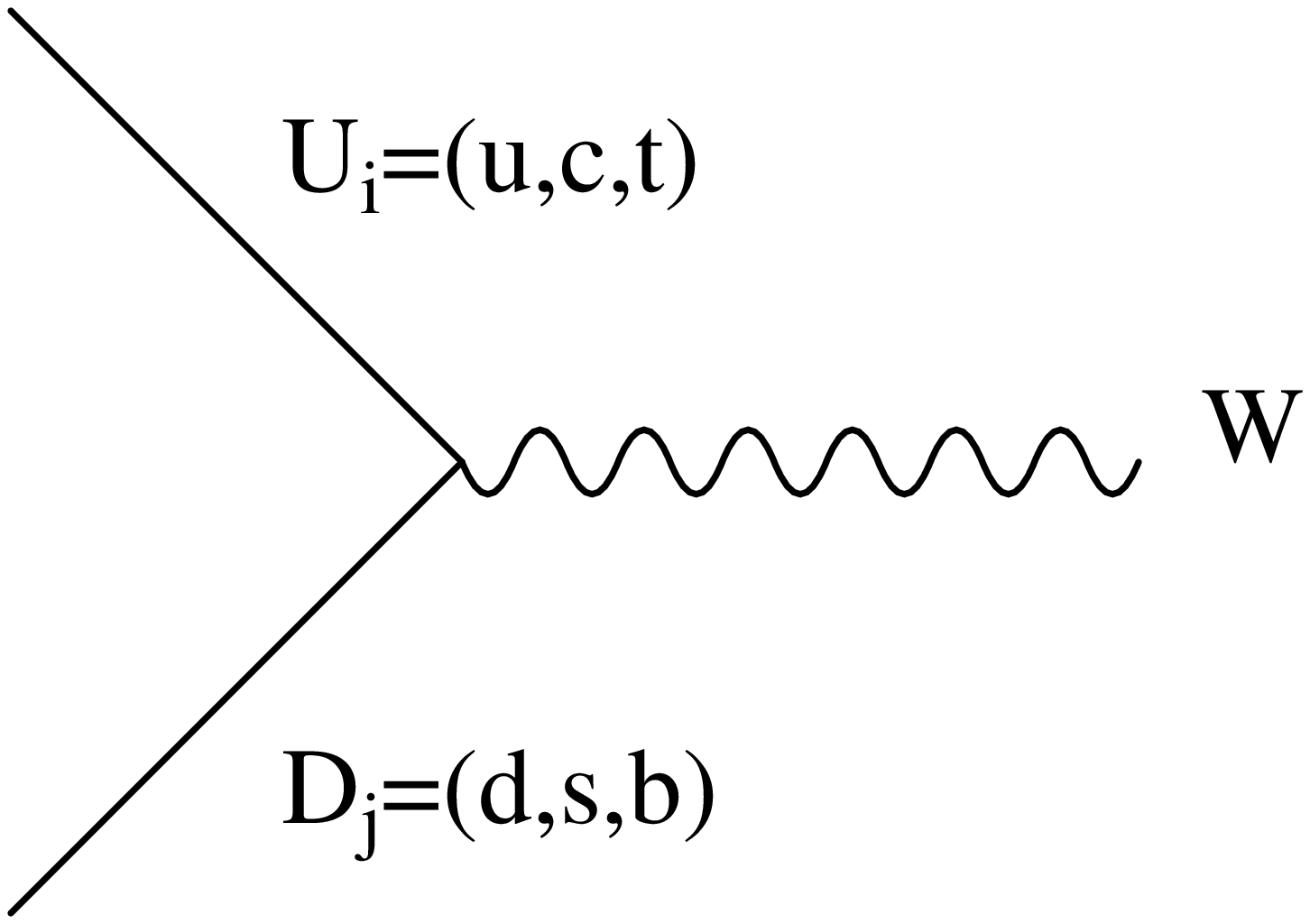,width=\linewidth}
\end{minipage}
\hspace{.2 cm} $\propto \; V_{ij}: VV^\dagger = 1$
\end{center}
\item[In the Lepton sector:]{The three charged leptons, $e$, $\mu$,
$\tau$, and the corresponding neutrinos have universal gauge
interactions, while the individual lepton numbers are exactly
conserved.}
\end{description}

\subsection{Unitarity of the CKM matrix}

In the quark sector, which is the subject of this talk, it remains
true that the numerically most precise test of the unitarity of
the CKM matrix comes from the sum of the squares of the first row.
Using current PDG numbers, one has
\begin{align}
&\quad |V_{ud}|^2 & \ngs \, +& \quad |V_{us}|^2 & \ngs \, +
& \; |V_{ub}|^2 & \ngs \, =&\nonumber\\
&0.9477(16) & \ngs \,+& \, 0.0481(10) & \ngs \, + & \; 10^{-5} &
\ngs \, =& \; 0.9958(19) \nonumber
\end{align}
where the three individual contributions are indicated with the
respective uncertainties. Should the deviation from 1 of this number be
considered significant? I suspect not, since I rather think that the
errors on the first two entries, both theoretically dominated, may be
slightly underestimated. Which does not mean that any
possible clarification of this point would not be welcome.

\subsection{Genuine Flavour Changing Neutral Current
processes}

Needless to say, checking the unitarity of the CKM matrix is not
really the point, since $V_{ij}$ is unitary almost by definition.
The real underlying question is rather: Is $\bar U_i \, V_{ij} \,
\gamma_{\mu} \, D_j \, W_{\mu}$ the {\it only} source of flavour
breaking {\it and} CP violation at the weak scale? The answer to
this question brings to the screen what I would like to call {\it
genuine} Flavour Changing Neutral Current processes (FCNC):  a
process only induced, at short distances, by a calculable
electroweak loop. Such processes are not the whole story. Their
use in comparison with experiment may be severely limited by the
inability to compute reliably the relevant matrix elements (see
the $\epsilon' /\epsilon$ story). Furthermore, processes with some
long distance contribution may also bring significant information
(as is the case, e.g., for $ B\rightarrow K  \pi$\cite{Neubert}).
Nevertheless genuine FCNCs represent at present the main testing
ground for the FS of the SM. For this reason it is crucial to have
in mind the complete list of the genuine FCNC processes that have
been observed so far. Such list is given in Table~\ref{tab:1}.
\begin{table*}[!bt]
\begin{center}
\begin{tabular}{|c||c|c|}
\hline
& \raisebox{0pt}[16pt][8pt]{Exp} & \raisebox{0pt}[16pt][8pt]{Th} \\
\hline \hline \raisebox{0pt}[16pt][8pt]{$\epsilon$} &
\raisebox{0pt}[16pt][8pt]{$(2.271\pm 0.017) 10^{-3}$} &
\raisebox{0pt}[16pt][8pt]{$\propto \eta (A-\rho)$} \\
\hline \raisebox{0pt}[16pt][8pt]{$\epsilon'/\epsilon $} &
\raisebox{0pt}[16pt][8pt]{$(17.2\pm 1.8) 10^{-4}$} &
\raisebox{0pt}[16pt][8pt]{$(1\div 30) 10^{-4}$} \\
\hline \raisebox{0pt}[16pt][8pt]{$BR(B\rightarrow \chi_s \gamma)$}
& \raisebox{0pt}[16pt][8pt]{$(3.22\pm 0.40) 10^{-4}$} &
\raisebox{0pt}[16pt][8pt]{$(3.50\pm 0.50) 10^{-4} $}\\
\hline \raisebox{0pt}[16pt][8pt]{$\Delta m_{B_d}$} &
\raisebox{0pt}[16pt][8pt]{$(0.487\pm 0.014)ps^{-1}$} &
\raisebox{0pt}[16pt][8pt]{$\propto (1-\rho)^2+\eta^2$}\\
\hline \raisebox{0pt}[16pt][8pt]{$A(B_d\rightarrow J/ \Psi K_S)$}
& \raisebox{0pt}[16pt][8pt]{$0.61 \pm 0.12$} &
\raisebox{0pt}[16pt][8pt]{$\frac{2 \eta (1-\rho)}{(1-\rho)^2 + \eta^2}$}\\
\hline \raisebox{0pt}[16pt][8pt]{$[BR(K^+\rightarrow \pi^+ \nu
\bar \nu)$} & \raisebox{0pt}[16pt][8pt]{$(1.5^{+3.4}_{-1.2} )
10^{-10}$} &
\raisebox{0pt}[16pt][8pt]{$(0.8 \pm 0.3) 10^{-10}]$}\\
\hline \raisebox{0pt}[16pt][8pt]{$[\frac {\Delta m_{B_s}}{\Delta
m_{B_d}}$} & \raisebox{0pt}[16pt][8pt]{$\geq 30(95\%C.L.)$} &
\raisebox{0pt}[16pt][8pt]{$\propto [(1-\rho)^2+\eta^2]^{-1}$]}\\
\hline
\end{tabular}
\caption{} \label{tab:1} \end{center}
\end{table*}

Other than its still rather limited number of entries, a few
things have to be remarked in this Table.

The only entry which allows at present a direct and numerically
significant comparison between experiment (second column) and
theory (third column), using independent information on the CKM
matrix, is provided by the branching ratio $BR(B\rightarrow \chi_s
\gamma)$\cite{exp-btosg}. The agreement is very satisfactory. An
important effort in the relevant theoretical calculation has been
made in recent years\cite{th-btosg}. Nevertheless I believe that a
$15\%$ error on the theoretical prediction is still there, at list
conservatively.

The information \cite{eps-ex} on the second row about
$\epsilon'/\epsilon$ is conceptually not less
significant, as remarked below. But the still large theoretical
uncertainties from the relevant matrix elements, more than one,
prohibit a stringent numerical test\cite{eps-th}.

\begin{figure*}[!bht]
\begin{center}
\centerline{\epsfig{file=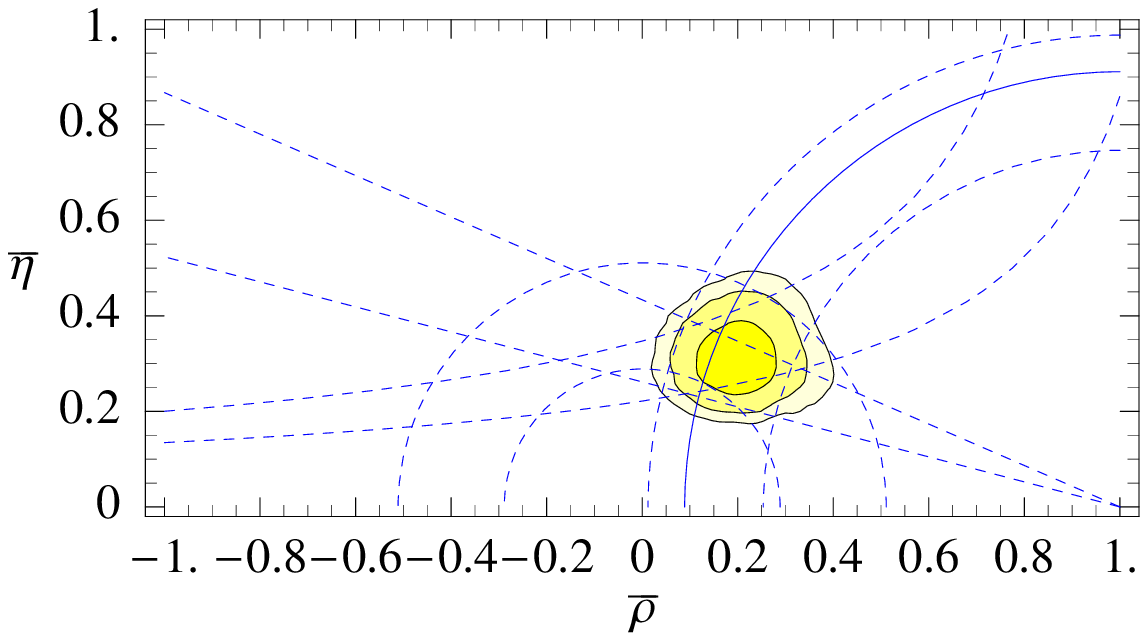,width=  0.6 \linewidth,
height=7.5 truecm}} \caption{} \label{fig:2}
\end{center}
\end{figure*}

As well known, the CP violating parameter $\epsilon$, as the
mixing $\Delta m_{B_d}$ and the CP asymmetry $A(B_d\rightarrow J/
\Psi K_S)$ serve to determine the Wolfenstein parameters $\rho$
and $\eta$, together with the last entry of Table 1, the limit on
$\frac {\Delta m_{B_s}}{\Delta m_{B_d}}$. The experimental number
on the $B_d$ asymmetry comes from BABAR\cite{BABAR asym} and does
not include the (higher) BELLE result\cite{BELLE-asym}, presented
during the Conference. The possibility of making a satisfactory
fit of these data, including the (dominant) theoretical
uncertainties, is again non trivial and important. The result of
my own fit is shown in Fig.~\ref{fig:2}. The contours are at $68$,
$95$, $99\%$ C.L. respectively.
The agreement in the $\rho-\eta$ plane is certainly significant.
At the present state of knowledge, I would summarize it by saying
that:
\begin{enumerate}
\item[i)]{no sign of new interactions exists so far in $\Delta B=2$ or
$\Delta S=2$;}
\item[ii)]{CP violation looks qualitatively or
semi-quantitatively as expected in the SM.}
\end{enumerate}
The possibility that the SM phase $\delta(CKM)$ be equal to zero,
with CP violation accounted for by non SM physics, is, for the
first time, rather strongly disfavoured.

\section{CP violation as of 2001}\label{sec:CP}

\subsection{A qualitative change}

Although technically correct, to say that CP violation looks as expected in
the SM is an understatement. The situation in CP violation has very clearly
changed in the past two years. This change can be summarized by making
reference to a general effective Lagrangian accounting for CP violation

\begin{equation}
  {\cal L}^{CP} ={\cal L}^{\Delta F =0} + {\cal L}^{\Delta F =1} + {\cal
L}^{\Delta F =2}.
\end{equation}

On the $\Delta F =0$ piece we still have only limits, a new
preliminary one on the electron Electric Dipole Moment, $d_e< 1.5
\cdot 10^{-27} \, \textrm{e cm}$, from Cummings and
Regan\cite{Cummings}, and on the neutron EDM,  $d_N< 6.3
\cdot10^{-26} \, \textrm{e cm}$. It is more and more true, in my
view, that searching for EDMs is a worthy enterprise.

The big change, however, has occurred in the flavour violating
pieces of ${\cal L}^{CP}$. We now know for sure that CP violation
exists in three different channels, $\Delta S =1$ (from
$\epsilon'/\epsilon$), $\Delta S =2$ (from $\epsilon$) and $\Delta
B =2$ (from $A(B_d\rightarrow J/ \Psi K_S)$). Experimental hints
existed already before, but these are now established facts.
Looking back at the history of CP violation, there is little doubt
that this is almost a revolution.

\subsection{A general parametrization of CP violation in $\Delta F =2$}

It is not difficult to conceive of motivated extensions of the SM
which, while accounting for the observations so far, may introduce
large deviations from the SM in other observables\cite{masiero}.
In view of this, it makes sense to consider a general
parametrization of CP violation in $\Delta F =2$. Concentrating
for the moment on the down quark sector, the CP violating
effective Lagrangian in the $\Delta F =2$ sector can be written at
short distances in the SM as
\begin{equation}
{\cal L}^{\Delta F=2} = \sum_{i,j=d,s,b} (V_{td_i}V_{td_j}^*)^2 \,
{\cal  C} \, (\bar d_i \gamma_{\mu} (1 - \gamma_5) d_j)^2
\end{equation}
where ${\cal  C}$ is an overall real coefficient originating from a top loop.
Hence ${\cal  C}$ is known. A small deviation from the above equation,
affecting $\epsilon$, can be easily accounted for and is not relevant to the
present discussion.

A general form of ${\cal L}^{\Delta F=2}$ can, on the other hand,
be written as
\begin{equation}
{\cal L}^{\Delta F=2} = \sum_{\alpha} \sum_{i,j=d,s,b}
(V_{td_i}V_{td_j}^*)^2 \, {\cal  C} _{ij}^{\alpha} \;
O_{ij}^{\alpha}
\end{equation}
where $O_{ij}^{\alpha}$ stand for the 4 fermion operators with
different Lorentz structures and given flavour indices $i, j$. At
the same time the ${\cal C}_{ij}^{\alpha}$ are complex numerical
coefficients, normalized for matter of convenience to
$(V_{td_i}V_{td_j}^*)^2$. In principle one would like to measure
not only the matrix elements $V_{td_i}$ for the different $i$,
but also the ${\cal C}_{ij}^{\alpha}$ and eventually compare with
the SM form. This task has to be confronted with the experimental
information in principle accessible: the determination of the
three complex matrix elements
\begin{equation}
\langle K| {\cal L}^{\Delta F}| \bar K \rangle, \; \langle B_d|
{\cal L}^{\Delta F}| \bar B_d \rangle, \; \langle B_s| {\cal
L}^{\Delta F}| \bar B_s \rangle.
\end{equation}

To guide this comparison, a useful progression of hypotheses is
made, i.e.:
\begin{enumerate}
\item{Minimal Flavour Violation (MFV)\cite{MFV}: As in the SM one takes
only one 4F-operator, $\alpha=LL$, but one leaves free the real,
flavour independent coefficient in front of it. In this case only
one new parameter is introduced. As easily seen, neither
$A(B_d\rightarrow J/ \Psi K_S)$ nor  $\frac{\Delta m_{B_s}}{\Delta
m_{B_d}}$ can possibly deviate from their SM values.}
\item{Generalized MFM\cite{GMFV}: More than one operator can now
contribute to ${\cal L}^{\Delta F=2}$, but the ${\cal
C}_{ij}^{\alpha}$ are taken to be real and $i,j$-indep. In this
case the moduli of the three measurable matrix elements become
free. Hence only $A(B_d\rightarrow J/ \Psi K_S)$ cannot deviate
from the SM value.}
\end{enumerate}

This progression of hypotheses can be useful for a practical
orientation of the analysis, in case of problems. The real
situation, if problems indeed occurred, may be different, however.
Deviations from the SM could for example also enter in ${\cal
L}^{\Delta F=1}$. I emphasize, quite in general, that there is
still ample room for deviations from the SM. To the purpose of
making them evident, it will be useful to consider, among the
possible tests, a series of possible ``null'' experiments. In the
SM a number of CP asymmetries should either be equal to each other
or to zero, as listed below:
\begin{align}
&A(B_d \rightarrow J/\Psi K_S) \; & \ngs \simeq &\; \; A(B_d
\rightarrow \phi K_S) && \nonumber \\
&A(B_d \rightarrow J/\Psi K_S) \; & \ngs \simeq &\; \;A(B_d \rightarrow
\pi^0 K_S) &&
\nonumber\\ &A(B^{\pm} \rightarrow \phi K^{\pm}) & \ngs \simeq & \; \;
A(B^{\pm}
\rightarrow \pi^{\pm} K^0) & \ngs \simeq &\; \;0\nonumber\\
&A(B_d \rightarrow \chi_s \gamma) & \ngs \simeq & \; \; 0; \;
A(B_s \rightarrow J/\Psi \phi) & \ngs \simeq &\; \;0\nonumber
\end{align}
The various ``$\simeq$'' signs do not have equal validity, as it
has been discussed in the different cases. These relations can be
taken however as a fair first approximation.

For lack of time I do not discuss possible CP violation in the
charm quark sector. I only notice that some preliminary indications in
charm mixing, reported at previous conferences \cite{charm}, have not
been apparently confirmed here\cite{Roudeau}.

\section{A hidden deeper structure?}

Having discussed the present evidence for the empirical adequacy
of the FS of the SM, I now turn to the more difficult problem of
trying to see if a deeper structure is possibly hidden in the
pattern of quark masses and mixings. As I said at the beginning,
the difficulty of this problem in no way is an excuse for not
trying to attack it.

As well known the problem is hard also because the basic
parameters decribing flavour in the SM, the Higgs Yukawa
couplings, are not all experimentally accessible, even in
principle: while knowledge of the Yukawa matrices $\lambda_U,
\lambda_D$ in the quark sector determines the masses and mixings,
$m_U, m_D$ and $ V_{CKM}$, the reverse is not true. There is
qualitative evidence for a hierarchical structure of the type
\begin{eqnarray}
\lambda_{ij}=c_{ij}\, x_i\, x_j & \,  \,  \text{ \small with}\,
\,x_3 \gg x_2 \gg x_1 \nonumber \\ & \! \! \text{\small and }\,
c_{ij}=O(1)
\end{eqnarray}
but to go further is not easy, to say the least. Although several
conceptually different approaches have been attempted even in the
last two years (Anarchy\cite{Anarchy},
Extra-dimensions\cite{Extra-D}, RG flows\cite{RG-flows}, etc.),
time forces me to limit my attention to a more conventional
approach, based on the so called ``Textures''. I make this choice
because the new data discussed above allow for the first time a
significant quantitative comparison with the expectations in this
framework.

\subsection{Textures defined}

It may be useful to define as precisely as possible what I mean by
``Texture''. There is some flavour basis, at some scale, in which:
\begin{enumerate}
\item{some of the coefficients $c_{ij}$ vanish;}
\item{for some (or all the)  $c_{ij}$, $|c_{ij}|=|c_{ji}|$;}
\item{no other precise relation is allowed among the different
$c_{ij}$;}
\item{when the Yukawa matrices for the up and down quarks are diagonalized, the
physical masses and angles do not result from accidental
cancellations between the various partial contributions involving
the $c_{ij}$.}
\end{enumerate}

The question that we ask is whether there are relations, exact or
approximate, between the physical observables (masses and angles)
implied by some texture, as defined above. An important
qualification, implicit in 3 and 4, is that these relations should
not change in a significant way upon variations of order unity of
the $c_{ij}$ (with 1 and 2 above satisfied).

This is a simple-minded approach. Even if one finds such relations, we may
not be able to interpret their meaning. I shall come back to this point. It is
nevertheless worth a try.

\subsection{Texture predictions}

In a way or another, many have tried to answer the question raised
in the previous subsection. I think that there is in fact a unique
set of relations, coherently obtainable following these rules,
which are not manifestly inconsistent with the presently known
data. They are\cite{textures}
\begin{eqnarray}
&&\left|V_{us}\right| = 
\left|\sqrt{\frac{m_d}{m_s}}-e^{i\phi}\sqrt{\frac{m_u}{m_c}}\right|\\
&&\left|\frac {V_{ub}}{V_{cb}}\right| = \sqrt{\frac{m_u}{m_c}}\\
&&\left|\frac {V_{td}}{V_{ts}}\right| = \sqrt{\frac{m_d}{m_s}}
\end{eqnarray}
where $\phi$ is an arbitrary phase and the various masses are
meant to be taken at the same scale, above or at the charm quark
mass. The unique texture which produces them\cite{unique texture}
is
\begin{displaymath}
\lambda_{U, \, D} = \lambda_{3, \, 3}^{U, \, D} \; \begin{pmatrix}
0 & \epsilon^{\prime} &  0\\[0.5cm]
  \epsilon^{\prime} & \epsilon & O(\epsilon) \\[0.5cm]
  0 &  O(\epsilon) & 1 \end{pmatrix}
\end{displaymath}
with the small parameters $\epsilon^{\prime} \ll \epsilon \ll 1$ being
U,D-dependent. The phases of the various entries are not indicated
because they are irrelevant. It is important, on the contrary,
that $|\lambda_{12}|=|\lambda_{21}|$. Only the size of the 23 and
32 entries matters, as I will show, but not their precise values. In
particular they do not need to be symmetric.

By a perturbative diagonalization of these Yukawa matrices the
above relations are readily obtained up to corrections of relative
order $ \epsilon$, i.e. a few $\%$. The explicit diagonalization
also shows that the phase $\phi$ in $V_{us}$ is approximately
equal to the angle $\alpha$ of the usual unitarity
triangle\cite{alpha}.

The texture zeros are crucial. One may in fact wonder how small
should the corresponding entries be to maintain the texture
relations in an approximate sense. Insisting on relative
corrections not exceeding $O(\epsilon)$, it is readily seen that
$\lambda_{11} \leq (\epsilon^{\prime})^2$, $\lambda_{13} \leq
\epsilon \epsilon^{\prime}$ and $\lambda_{31} \leq
\epsilon^{\prime}$.

\subsection{Comparison with data}

To compare the texture relations with data I use\cite{mass-ratios}
\begin{align}
& Q=\frac{\frac {m_s}{m_d}}{\sqrt{1- (\frac {m_u}{m_d})^2}}=22.7
\pm 0.8, \\ & \frac {m_c}{m_s} =9.3 \pm 3.0,
\end{align}
and, but this is a less important entry,
\begin{equation}
\frac {m_u}{m_d}=0.533 \pm 0.043.
\end{equation}

The combination $Q$, rather the individual ratios $\frac
{m_s}{m_d}$ and $\frac {m_u}{m_d}$, is determined without extra
assumptions using chiral perturbation theory. The somewhat
conservative error on $\frac {m_c}{m_s}$ is not, in my view,
without justification.

\begin{figure*}[!hbt]
\begin{center}
\begin{minipage}[c]{9.2cm}
\epsfig{file=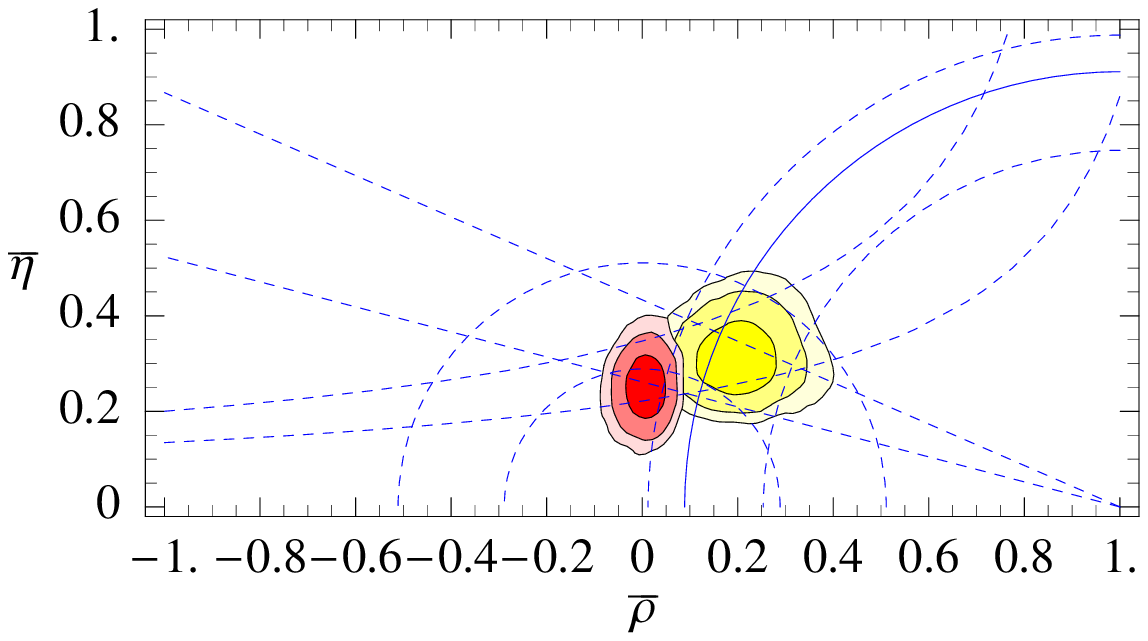,width=9.0truecm,height=7.0truecm}
\end{minipage} {a}
\vspace{.5cm}
\begin{minipage}[c]{9.2cm}
\epsfig{file=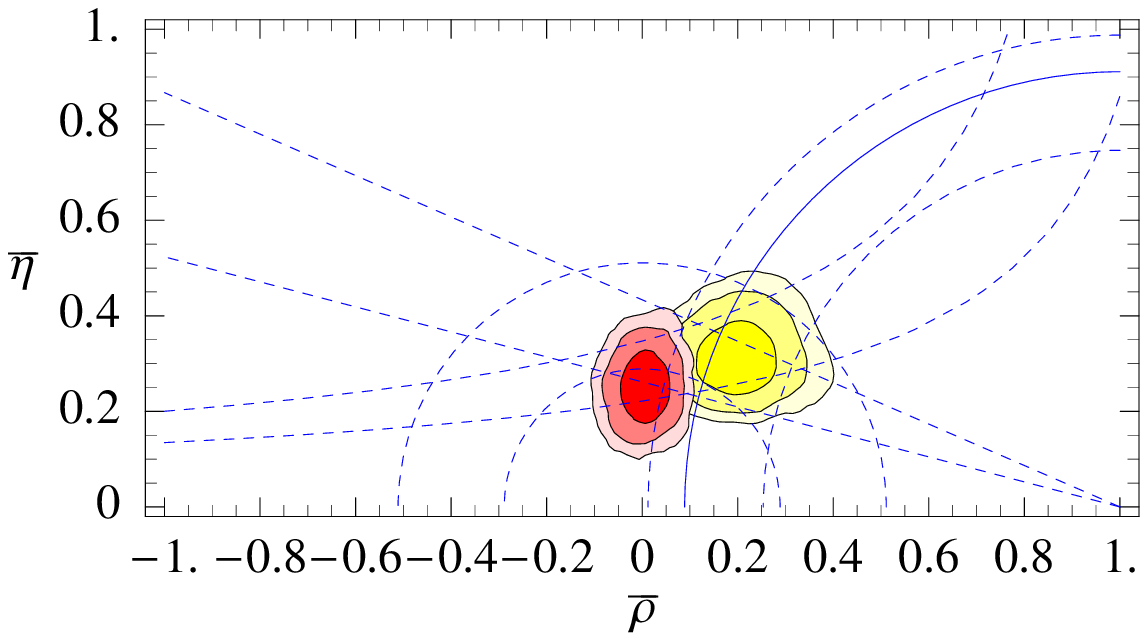,width=9.0truecm,height=7.0truecm}
\end{minipage} {b}
\end{center}
\caption{} \label{fig:3}
\end{figure*}

With this information on the quark mass ratios, the texture
relations allow to determine the Wolfenstein parameters $\rho$ and
$\eta$, as shown by the smaller contours in Fig.s~\ref{fig:3}. As
before the contours are at $68$, $95$, $99\%$ C.L. respectively.
The difference between Fig.~\mbox{\ref{fig:3}a} and
\mbox{\ref{fig:3}b} is that in \mbox{\ref{fig:3}b} the
coefficients $c_{23}$ and $c_{32}$ are allowed to vary at random
and independently from $1/3$ to $3$.

This determination of $\rho$ and $\eta$ can be compared with
experiment, assuming no surprise in genuine FCNCs with respect to
SM expectations \cite{BHR}. To this purpose it is enough to superimpose
the contours as obtained from the textures to the fit of the data
discussed in Section 2. This
comparison is also shown in Fig.s~\ref{fig:3}.

As anticipated, this comparison starts to be significant. To judge
the result, it may be useful to wait for a settlement of the data.
Some tension is appearing, however, which would be enhanced if the
BELLE result on $A(B_d\rightarrow J/ \Psi K_S)$ were included. Two
problems seem to emerge: the limit on $\frac{\Delta
m_{B_s}}{\Delta m_{B_d}}$ is too high, as it also seems to be the
case for $\left|\frac{V_{ub}}{V_{cb}}\right|$.

In case these difficulties were confirmed, ways out that are
proposed involve a modification of the texture\cite{RRR}, like
\begin{eqnarray}
\frac {\lambda_{U,D}}{ \lambda_{3,3}^{U,D}} \;\; = &
\begin{pmatrix}
   0 &\epsilon^{\prime} & 0\\
\epsilon^{\prime} &  0 & O(\epsilon) \\
   0 &  O(1) &  1 \end{pmatrix} \nonumber\\
  \textrm{or} \;
& \begin{pmatrix}
   0 &\epsilon^{\prime} &  O(\sqrt{\epsilon} \epsilon \prime) \\
\epsilon^{\prime} &  \epsilon & O(\epsilon) \\
    O(\sqrt{\epsilon} \epsilon \prime) &  O(\epsilon) & 1 \end{pmatrix}
\end{eqnarray}

  or the inclusion of some non
standard FCNC effects.

In the first case, the loss of predictivity has to be watched. To restore it
one may have to include leptons as well. The implication of the second option
is that   some other deviation in FCNCs should be seen.

\subsection{Where does the Texture come from?}

If the texture turned out to be successful, we would remain with
the problem of interpreting its origin or its meaning. The
interpretation I favour calls for the relevance of a $U(2)$
symmetry acting on the first 2 generations as a doublet,
supplemented with a straightforward spurion analysis\cite{U2}, \cite{BHR}.
The spurions may only include a doublet $\phi_a $, a triplet $S_{\{ a,b
\}}$ and an antisymmetric singlet $A_{[ a,b ]}$. The $U(2)$
symmetry breaking pattern would have to be as follows

\begin{equation*}
U(2) \, \stackrel{\epsilon: \, \phi_a , \, S_{\{ a,b \}}}{\longrightarrow}
\; U(1)
\stackrel{\epsilon^\prime: \, A_{[ a,b ]}}{\longrightarrow} \emptyset
\end{equation*}\\
since this would lead to
\begin{equation*}
\frac {\lambda_{U,D}}{ \lambda_{3,3}^{U,D}}\Rightarrow \begin{pmatrix} 0 &
A_{12} & 0 \\
    -A_{12} & S_{22} & \phi_2 \\
0 &  \phi_2 & 1 \end{pmatrix} \Rightarrow \begin{pmatrix} 0 &
\epsilon^\prime & 0
\\
    \epsilon^\prime  & \epsilon & O(\epsilon) \\
0 & O(\epsilon)   & 1 \end{pmatrix}
\end{equation*}

I admit, though, that it may be difficult to prove the relevance of this
$U(2)$ symmetry without additional experimental information on flavour
physics.

\section{Summary and conclusions}

The significant flavour measurements performed in the last two years are
compatible with the FS of the SM. It is clear, however, that the test is
still at a qualitative or only semi-quantitative level. Large deviations from
the expectations of the SM in many observables are still possible.

The status of CP violation has seen a qualitative change. $CP$
violation is established in $\Delta S=1$, $\Delta S=2$ and $\Delta
B=2$ channels. As a result, the possibility that the CKM phase
vanishes, with all of $ CP$ violation coming from non SM sources
looks, for the first time, clearly disfavoured.

Finally, as a concrete attempt to make sense of the data on quark
masses and mixings, a single texture emerges which can be
meaningfully compared with the data. It may even have something to
do with reality. Quite in general I think that theories of flavour
should aim at a comparable level of concreteness.

\section*{Acknowledgements}

I thank Andrea Romanino for his help in the fits of Fig. 2 and Fig. 3.
This work was supported by the ESF under the RTN contract
HPRN-CT-2000-00148.

\end{document}